\documentclass[aps,prd,eqsecnum,11pt,preprintnumbers,nofootinbib]{revtex4}
\usepackage{amssymb,amsmath,bm,color}
\usepackage{graphicx}
\usepackage{graphics}
\usepackage{float}
\pdfoutput=1
\newcommand{\bea}{\begin{eqnarray}}
\newcommand{\eea}{\end{eqnarray}}
\newcommand{\be}{\begin{equation}}
\newcommand{\ee}{\end{equation}}
\newcommand{\bes}{\begin{subequations}}
\newcommand{\ees}{\end{subequations}}
\newcommand{\nn}{\nonumber}
\def\lag{\langle}
\def\rag{\rangle}
\def\nbox#1#2{\vcenter{\hrule \hbox{\vrule height#2in
\kern#1in \vrule} \hrule}}
\def\sq{\,\raise.5pt\hbox{$\nbox{.09}{.09}$}\,}
\def\sqb{\,\raise.5pt\hbox{$\overline{\nbox{.09}{.09}}$}\,}

\begin{document}

\title{The Trace Anomaly and\\
 Dynamical Vacuum Energy in Cosmology}

\author{Emil Mottola}

\affiliation{Theoretical Division, Los Alamos National Laboratory\\
Los Alamos, NM 87545 USA\\
{\rm and}\\
Theoretical Physics Group, PH-TH, CERN\\
CH-1211, Geneva 23, Switzerland\\
E-mail: emil@lanl.gov \ {\rm and}\  emil.mottola@cern.ch\\ \\
{\rm Invited Talk at the Ninth Conference on}\\
Quantum Field Theory under the Influence of External Conditions\\
{\rm The University of Oklahoma, Norman, OK\\ September 21-25, 2009\\
To appear in the} Proceedings of QFEXT09\\}

\begin{abstract}
The trace anomaly of conformal matter implies the existence of massless scalar poles
in physical amplitudes involving the stress-energy tensor. These poles may be 
described by a local effective action with massless scalar fields, which couple to 
classical sources, contribute to gravitational scattering processes, and can have 
long range gravitational effects at macroscopic scales. In an effective field theory
approach, the effective action of the anomaly is an infrared relevant term that should 
be added to the Einstein-Hilbert action of classical General Relativity to take account 
of macroscopic quantum effects. The additional scalar degrees of freedom contained
in this effective action may be understood as responsible for both the Casimir effect
in flat spacetime and large quantum backreaction effects at the horizon scale of
cosmological spacetimes. These effects of the trace anomaly imply that the cosmological
vacuum energy is dynamical, and its value depends on macroscopic boundary conditions 
at the cosmological horizon scale, rather than sensitivity to the extreme ultraviolet
Planck scale.

\end{abstract}


\preprint{CERN-PH-TH/2010-043}
\preprint{LA-UR 10-01043}

\maketitle
\pagebreak

\section{The Cosmological Constant and Vacuum Energy}

In classical General Relativity, the requirement that the field eqs. 
involve no more than two derivatives of the metric tensor allows for the 
possible addition of a constant term, the cosmological term $\Lambda$, 
to Einstein's eqs.
\be
R^{\mu}_{\ \nu} - \frac{R}{2} \,\delta^{\mu}_{\ \nu} + \Lambda \,\delta^{\mu}_{\ \nu} 
= \frac{8\pi G}{c^4}\, T^{\mu}_{\ \nu}.
\label{Ein}
\ee
If transposed to the right side of this relation, the $\Lambda$ term corresponds 
to a constant energy density $\rho_{\Lambda} = c^4\Lambda /8\pi G$ and isotropic 
pressure $p_{\Lambda} = - c^4\Lambda/8\pi G$ permeating all of space uniformly, 
and independently of any localized matter sources. Hence, even if the matter 
stress tensor $T^\mu_{\ \nu} = 0$, a cosmological term causes spacetime to 
become curved with a radius of curvature of order $\vert\Lambda\vert^{-\frac{1}{2}}$.
In purely classical physics there is no natural scale for $\Lambda$. 
Indeed if $\hbar =0$ and $\Lambda = 0$, there is no fixed length scale 
at all in the vacuum Einstein equations, $G/c^4$ being simply a conversion
factor between the units of energy and those of length. Hence $\Lambda$ 
may take on any value whatsoever with no difficulty (and with no explanation) 
in classical General Relativity. 

As soon as we allow $\hbar \neq 0$, there is a quantity with the dimensions 
of length that can be formed from $\hbar, G$, and $c$, namely the Planck length,
\be
L_{pl} \equiv \left( \frac{\hbar G}{c^3}\right)^{\frac{1}{2}}
= 1.616 \times 10^{-33}\,{\rm cm}.
\label{Lpl}
\ee
Hence when quantum theory is considered in a general
relativistic setting, the quantity,
\be
\lambda \equiv \Lambda L_{pl}^2 = \frac{\hbar G \Lambda}{c^3}
\label{Lnum}
\ee
becomes a dimensionless pure number, whose value one might expect 
a theory of gravity incorporating quantum effects to address.

Some eighty years ago W. Pauli was apparently the first to consider the
question of the effects of quantum vacuum fluctuations on the the curvature 
of space \cite{Pau}. Pauli recognized that the sum of zero point energies of 
the two transverse electromagnetic field modes {\it in vacuo}
\be
\rho_{\Lambda} = 2 \int^{L_{min}^{-1}}\frac{d^3 {\vec  k}}{(2\pi)^3} 
\frac{\hbar \omega_k}{2}
 = \frac{1}{8\pi^2} \frac{\hbar c}{\ L_{min}^{\ 4}} = -p_{\Lambda}
\label{zeropt}
\ee
contribute to the stress-energy tensor of Einstein's theory as would an 
effective cosmological term $\Lambda > 0$. Since the integral (\ref{zeropt}) 
is quartically divergent, an ultraviolet cutoff $L_{min}^{-1}$ of (\ref{zeropt}) 
at large $|\vec k|$ is needed. Taking this short distance cutoff $L_{min}$ to be 
of the order of the classical electron radius $e^2/mc^2$, Pauli concluded that if 
this estimate were correct, Einstein's theory with this large a $\Lambda$ would 
lead to a universe so curved that its total size ``could not even reach to 
the moon." If instead of the classical electron radius, the apparently natural 
but much shorter length scale of $L_{min} \sim L_{pl}$ is used to cut off the 
frequency sum in (\ref{zeropt}), then the estimate for the cosmological term in 
Einstein's equations becomes vastly larger, and the entire universe would be 
limited in size to the microscopic scale of $L_{pl}$ (\ref{Lpl}) itself, in even 
more striking disagreement with observation.

Clearly the naive estimate of the contribution of short distance modes of
the electromagnetic field to the curvature of space, by using (\ref{zeropt}) 
as a source for Einstein's eqs. (\ref{Ein}) is not correct. The question is why.
Here the Casimir effect may have something to teach us. The vacuum zero point 
fluctuations being considered in (\ref{zeropt}) are the same ones that contribute 
to the Casimir effect, but this estimate of the scale of vacuum zero point energy, 
quartically dependent on a short distance cutoff $L_{min}$, is certainly {\it not} 
relevant for the effect observed in the laboratory \cite{Cas}. In calculations of the 
Casimir force between conductors, one subtracts the zero point energy of the 
electromagnetic field in an infinitely extended vacuum (with the conductors absent) 
from the modified zero point energies in the presence of the conductors. It is this 
{\it subtracted} zero point energy of the electromagnetic vacuum, depending upon 
the {\it boundary conditions} imposed by the conducting surfaces, which leads 
to experimentally well verified results for the force between the conductors. 

In the renormalization procedure the ultraviolet cutoff $L_{min}^{-1}$ drops out, 
and the distance scale of quantum fluctuations that determine the magnitude 
of the Casimir effect is not the microscopic classical electron radius, as in Pauli's 
original estimate, much less the even more microscopic Planck length $L_{pl}$, 
but rather the relatively {\it macroscopic} distance $d$ between the conducting 
boundary surfaces. The resulting subtracted energy density of the vacuum 
between the conductors is
\be
\rho_v = -\frac{\pi^2}{720}\, \frac{\hbar c}{d^4} \,.
\label{Casimir}
\ee
This energy density is of the opposite sign as (\ref{zeropt}), leading 
to an attractive force per unit area between the plates of 
$0.013$ dyne/cm$^2$ $(\mu m/d)^4$, a value which is both independent 
of the ultraviolet cutoff $L_{min}^{-1}$, and the microscopic details of 
the atomic constituents of the conductors. This is a clear indication, 
confirmed by experiment, that the {\it measurable} effects associated 
with vacuum fluctuations are {\it infrared} phenomena, dependent upon 
macroscopic boundary conditions, which have little or nothing to do 
with the extreme ultraviolet modes or cutoff of the integral in (\ref{zeropt}).

By the Principle of Equivalence, local short distance behavior in a mildly
curved spacetime is essentially equivalent to that in flat spacetime.
Hence on physical grounds we should not expect the ultraviolet 
cutoff dependence of (\ref{zeropt}) to affect the universe in the large
any more than it affects the force between metallic conductors in the 
laboratory. 

In the case of the Casimir effect a constant zero point energy of the
vacuum, no matter how large, does not affect the force between the plates.
In the case of cosmology it is usually taken for granted that any effects of
boundary conditions can be neglected.  It is not obvious then what should 
play the role of the conducting plates in determining the magnitude of 
$\rho_v$ in the universe, and the magnitude of any effect of quantum zero 
point energy on the curvature of space has remained unclear from Pauli's 
original estimate down to the present. In recent years this has evolved from 
a question of fundamental importance in theoretical physics to a central one 
of observational cosmology as well. Observations of type Ia supernovae at 
moderately large redshifts ($z\sim 0.5$ to $1$) have led to the conclusion 
that the Hubble expansion of the universe is {\it accelerating} \cite{SNI}. 
This is consistent also with microwave background measurements \cite{WMAP}.
According to Einstein's equations accelerated expansion is possible if and 
only if the energy density and pressure of the dominant component of the 
universe satisfy the inequality, 
\be
\rho + 3 p = \rho\  (1 + 3 w) < 0\,.
\label{accond}
\ee
A vacuum energy with $\rho > 0$ and $w\equiv p_v/\rho_v = -1$ leads to an 
accelerated expansion, a kind of ``repulsive" gravity in which the relativistic
effects of a negative pressure can overcome a positive energy density in
(\ref{accond}). Taken at face value, the observations imply that some $74\%$ 
of the energy in the universe is of this hitherto undetected $w=-1$ dark 
variety \cite{SNI,WMAP}. This leads to a non-zero inferred cosmological 
term in Einstein's equations of 
\be
\Lambda_{\rm meas} \simeq (0.74)\, \frac{3 H_0^2}{c^2} 
\simeq 1.4 \times 10^{-56}\ {\rm cm}^{-2}
\simeq  3.6 \times 10^{-122}\ \frac{c^3}{\hbar G}\,.
\label{cosmeas}
\ee
Here $H_0$ is the present value of the Hubble parameter, approximately 
$73\, {\rm km/sec/Mpc} \simeq 2.4 \times 10^{-18}\, {\rm sec}^{-1}$. Thus 
the value of the cosmological dark energy inferred from the SN Ia data in 
terms of Planck units, $L_{\rm pl}^{-2} = \frac{c^3}{\hbar G}$, gives the 
dimensionless number in (\ref{Lnum}) the extremely small but finite value,
\be
\lambda \simeq 3.6 \times 10^{-122}\,.
\label{lmeas}
\ee
Explaining the value of this smallest number in all of physics is the
basic form of the {\it cosmological constant problem}.

As we have already noted, if the universe were purely classical, $L_{pl}$ 
would vanish and $\Lambda$, like the overall size or total age of the universe, 
could take on any value whatsoever without any technical problem of naturalness. 
Likewise as the Casimir effect makes clear, if $G=0$ and there are also no 
boundary effects to be concerned with, then the cutoff dependent zero point 
energy of flat space (\ref{zeropt}) could simply be subtracted, with no observable 
consequences. A naturalness problem arises only when the effects of 
quantum zero point energy on the large scale curvature of spacetime are 
considered. This is a problem of the gravitational energy of the quantum 
vacuum or ground state of the system at {\it macroscopic} distance scales, 
very much greater than $L_{pl}$, when both $\hbar \neq 0$ and $G \neq 0$.

\section{Effective Field Theory and Anomalies}

The treatment of quantum effects at distances much larger than any ultraviolet 
cutoff is precisely the context in which effective field theory (EFT) techniques 
should be applicable. This means that we assume that we do not need to know 
every detail of physics at extremely short distance scales of $10^{-33}$ cm or 
even $10^{-13}$ cm in order to discuss cosmology at $10^{28}$ cm scales. 
In EFT one assumes some organizing principle or symmetry of low energy 
dynamics, expresses degrees of freedom in terms of local fields having 
well-defined covariant transformation properties under the symmetry, and 
expands the effective action in local invariants of increasing number of 
derivatives of the fields. The dimensionful parameters multiplying the terms 
in the action determine the scale at which the derivative expansion is expected 
to break down.

The organizing principle in gravity is the Principle of Equivalence, {\it i.e.} invariance 
under general coordinate transformations, which greatly restricts the form of any 
EFT of gravity. In his search for field equations (\ref{Ein}) for a metric theory with 
universal matter couplings, which incorporates the Equivalence Principle 
automatically, but which is no higher than second order in derivatives of the 
metric, Einstein was using what we would now recognize as EFT reasoning. 
In an EFT quantum effects and any ultraviolet (UV) divergences they generate 
at very short distance scales are absorbed into a few, finite low energy effective 
parameters, such as $G$ and $\Lambda$. 

In extending Einstein's classical theory to take account of the quantum properties
of matter, the classical stress-energy tensor of matter $T^{\mu}_{\ \nu}$ becomes a
quantum operator, with an expectation value $\lag T^{\mu}_{\ \nu}\rag$. In this
{\it semi-classical} theory with both $\hbar$ and $G$ different from zero,
quantum zero-point and vacuum energy effects first appear, while the
spacetime geometry can still be treated classically. This is clearly an
approximation to a more exact treatment, which can be formally justified
by taking the number $N$ of matter degrees of freedom to infinity, and in
quantum states which are sharply peaked about their mean value.

Since the expectation value $\lag T^{\mu}_{\ \nu}\rag$ suffers from the 
quartic divergence (\ref{zeropt}), a regularization and renormalization
procedure is necessary in order to define the semi-classical EFT.
General coordinate invariance requires a more careful renormalization 
procedure than the simple subtraction of (\ref{zeropt}) which suffices for 
the original Casimir calculations in flat space. The UV divergent terms 
of the stress tensor contain subleading quadratic and logarithmic 
dependence upon the cutoff $L_{min}$ which must be isolated and  
removed in a way consistent with the Equivalence Principle to extract 
physical effects correctly. These more general renormalization procedures, 
involving {\it e.g.} proper time, covariant point splitting or dimensional 
regularization have been developed in the context of quantum field theory 
in curved spacetime \cite{BirDav}. The non-renormalizability of the classical 
Einstein theory poses no particular obstacle for this semi-classical EFT 
approach. It requires only that certain additional terms be added to the 
effective action to take account of UV divergences which are not of the 
form of a renormalization of $G$ or $\Lambda$. One such set of terms
that arise from a consistent covariant renormalization scheme are those
associated with the {\it trace anomaly} of $\lag T^{\mu}_{\ \nu}\rag$. The result 
of the renormalization program for quantum fields and their vacuum energy 
in curved space is that General Relativity can be viewed as a low energy 
quantum EFT of gravity, provided that the classical Einstein-Hilbert classical action 
is augmented by the additional terms required by the trace anomaly when 
$\hbar \neq 0$.

The essential physical assumption in any EFT approach is the hypothesis
of {\it decoupling}, namely that low energy physics is independent of very 
short distance degrees of freedom and the details of their interactions. All 
of the effects of these short distance degrees of freedom are subsumed into
a few phenomenological coefficients of the infrared relevant 
terms of the EFT. Notice that this will not be the case if the low energy 
$\Lambda$ relevant for dark energy and cosmology depends upon the 
quantum zero point energies of all fields up to some UV cutoff, as in 
(\ref{zeropt}). Taken seriously this would indicate {\it quartic} power sensitivity
of extreme infrared physics to the ultraviolet cutoff. In addition to violating
any intuitive notion of decoupling, this is clearly not how the Casimir effect
works. The hierarchy between the scale of electroweak symmetry breaking 
in the Standard Model compared to the Planck scale also suggests that low 
energy physics does not have even quadratic power law sensitivity 
to the extreme UV cutoff scale $L_{Pl}$.

Power law sensitivity to ultraviolet cutoffs may well be an artificial problem 
of a poor regularization technique, since for example it does not occur in 
dimensional regularization. On the other hand {\it logarithmic} scale sensitivity 
is the basis of renormalization group analyses, and in the case of the Standard 
Model has been verified experimentally \cite{L3}. The argument of a logarithm 
necessarily involves the ratio of a UV scale to an IR scale. This distinguishes 
logarithms from simple additive UV contributions to dimensionful quantities 
such as the Higgs mass, or the cosomological vacuum energy in (\ref{zeropt}). 
Logarithmic corrections to classical gravity arises from the conformal 
or trace anomaly \cite{CapDuffDes,BirDav}. 

Anomalies violate strict decoupling of UV from IR degrees of freedom in the sense 
that the effective action that describes them is necessarily non-local in terms of the 
original local field degrees of freedom. Because it is non-local in terms of the 
original fields, the usual EFT approach of expanding in local invariants with higher 
numbers of derivatives of those fields will miss the anomaly. Instead the non-local 
effective action of the anomaly must be added explicitly to the local EFT action. 
Alternatively, an anomaly generally implies massless poles signifying additional 
massless degrees of freedom which do not decouple, and these new degrees of 
freedom need to be added to the action to complete the low energy effective theory.

\section{The Axial Anomaly and Its Massless Pole}

The best known example of a quantum anomaly is the chiral or axial anomaly of QED, 
\cite{AdlBelJac} also present in QCD. Despite this there are some features
of the QED axial anomaly that remain somewhat underappreciated, which are
directly relevant to the gravity case. Therefore it is well to review the IR features
of the QED axial anomaly before proceeding to gravitational applications.

In QED the Dirac equation
\be
-i \gamma^{\mu} ( \partial_{\mu} - ieA_{\mu})\psi + m \psi = 0\,,
\label{Dirac}
\ee
implies that the vector current $J^{\mu} = \bar\psi \gamma^{\mu} \psi$ is conserved:
\be
\partial_{\mu} J^{\mu} = 0\,.
\label{vecons}
\ee
The axial current $J_5^{\mu} = \bar\psi\gamma^{\mu} \gamma^5 \psi$ (with
$\gamma^5 \equiv i\gamma^0 \gamma^1 \gamma^2 \gamma^3$) apparently obeys 
\be
\hspace{1cm} \partial_{\mu} J_5^{\mu} = 2 i m\, \bar\psi\gamma^5 \psi \qquad {\rm (classically)}.
\label{axclass}
\ee
In the limit of vanishing fermion mass $m\rightarrow 0$, the classical Lagrangian has a
$U_{ch}(1)$ global symmetry under $\psi \rightarrow e^{i\alpha\gamma^5}\psi$, in 
addition to $U(1)$ local gauge invariance, and $J_5^{\mu}$ is the Noether current
corresponding to this chiral symmetry. As is well known, both symmetries cannot be 
maintained simultaneously at the quantum level \cite{AdlBelJac}. Let us denote by 
$\lag J_5^{\mu}(z) \rag_{_A}$ the expectation value of the chiral current in the 
presence of a background electromagnetic  potential $A_{\mu}$. Enforcing $U(1)$ 
gauge invariance (\ref{vecons}) on the full quantum theory leads necessarily to a finite
axial current anomaly,
\be
\partial_{\mu}\lag J_5^{\mu}\rag_{_A}\Big\vert_{m=0} = \frac{e^2}{16\pi^2} \
\epsilon^{\mu\nu\rho\sigma}F_{\mu\nu}F_{\rho\sigma} = \frac{e^2}{2\pi^2}\,{\vec  E \cdot \vec B}\,,
\label{axanom}
\ee
in an external electromagnetic field.  Varying this expression twice with respect to the
external $A$ field and Fourier transforming, we see that the anomaly must appear in the amplitude,
\bea
&&\Gamma^{\mu \alpha\beta}(p,q) \equiv - i \int d^4x\int d^4y\, e^{ip\cdot x+ i q\cdot y}\  
\frac{\delta^2 \lag J_5^{\mu}(0) \rag_{_A}}{\delta A_{\alpha}(x) \delta A_{\beta}(y)} 
\nonumber\Bigg\vert_{A=0}\\
&&= ie^2 \int d^4 x \int d^4 y\, e^{i p\cdot x + i q \cdot y}\ \lag {\cal T}
J_5^{\mu}(0) J^{\alpha}(x) J^{\beta}(y)\rag\big\vert_{A=0}\,.
\label{GJJJ}
\eea
At the lowest one-loop order it is given by a triangle diagram with the axial
current $J_5^{\mu}$ at one vertex with four-momentum $k^{\mu}$, and the vector 
currents $J^{\alpha}$ and $J^{\beta}$ at the other two vertices, coupling to
photons with four-momenta $p^{\mu}$ and $q^{\mu}$ respectively. Momentum
conservation requires $k^{\mu} = p^{\mu} + q^{\mu}$.

\begin{figure}[htp]
\includegraphics[width=40cm, viewport=80 550 1000 660,clip]{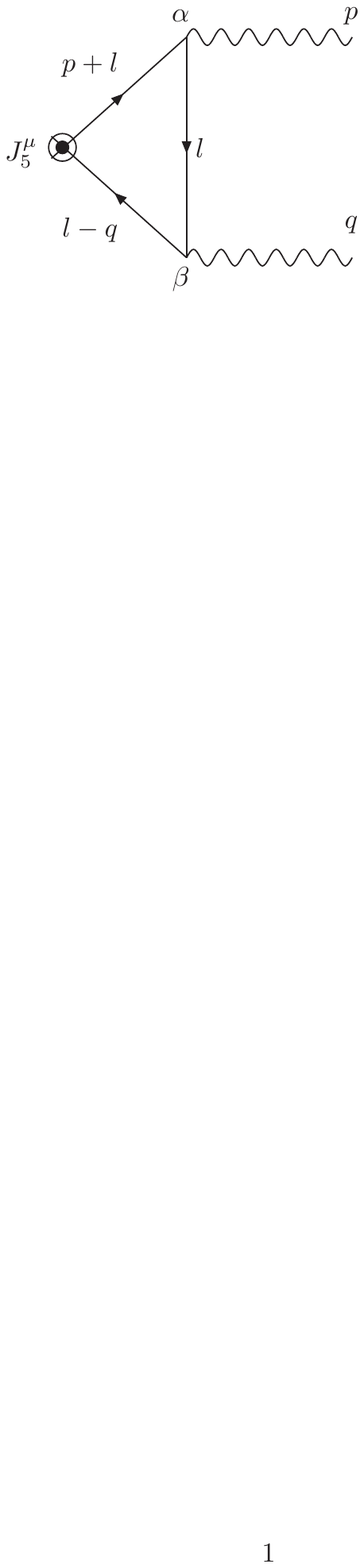}
\caption{The Axial Anomaly Triangle Diagram. The momentum integration variable is $l$.}
\label{Fig:tri}
\end{figure}

The anomaly may be regarded as a clash of symmetries. Either but not both of the two
classically valid relations (\ref{vecons}) and (\ref{axclass}) can be maintained at
the quantum level. There are various ways to see this, but intuitively, defining
the quantum amplitude (\ref{GJJJ}) at one-loop order introduces implicitly or
explicitly an additional scale into the amplitude, which violates naive identities 
dependent upon global scale and chiral invariance. Although this is usually
presented as a UV scale, necessary to regulate the triangle amplitude,
in fact it is not difficult to show that the axial anomaly (\ref{axanom}) is
{determined by its UV {\it finite}, {\it i.e.} cutoff independent  parts, 
together with Lorentz invariance, vector current conservation (\ref{vecons}), 
and Bose symmetry of exchanging the two external photon lines 
\cite{DolZak,Horejb,GiaMot}. Thus, the anomaly can be equally well thought 
of as due to the sensitivity of the amplitude (\ref{GJJJ}) to an {\it infrared} 
cutoff at large distances when $m=0$ and $k = p+ q = 0$.

Consistent with this infrared interpretation of the axial anomaly, one can
demonstrate explicitly the existence of a massless pole in the amplitude (\ref{GJJJ}),
and in the matrix element of the axial current between the vacuum
and the physical two-photon state, $\vert p,q\rangle$, giving \cite{GiaMot}
\bea
&&\lag 0 \vert J_5^{\mu}(0) \vert p,q\rag = i\Gamma^{\mu\alpha\beta}(p,q)
\tilde A_{\alpha}(p) \tilde A_{\beta}(q)\big\vert_{m^2=p^2=q^2=0}\nn\\
&& \qquad \qquad = \ \frac{ie^2}{2\pi^2 k^2}\ k^{\mu} \epsilon^{\alpha\beta \rho\sigma} p_{\rho}q_{\sigma}\tilde A_{\alpha}(p) \tilde A_{\beta}(q)\,,
\label{matpol}
\eea
When contracted with $k_{\mu}$ the pole cancels and we recover (\ref{axanom})
in momentum space. The existence of the pole at $k^2 = 0$ in a physical amplitude 
(\ref{matpol}) implies the existence of a propagating massless pseudoscalar $0^-$ 
intermediate state, with low energy long range effects. The discontinuity of the 
amplitude contains a $\delta$ function in the corresponding spectral representation,
the signal of a massless intermediate state \cite{Horejb,GiaMot}. 
This new state appears in the two-particle correlations of massless 
fermions moving collinearly at the speed of light, and are a genuine 
quantum effect. The effective action for the anomaly is non-local in 
terms of the original fields but can be rendered local by the introduction 
of two new massless pseudoscalar fields whose propagator gives
rise to the $k^{-2}$ pole in (\ref{matpol}).  

In real QED these infrared effects are suppressed by the non-zero physical 
electron mass $ m> 0$, and the additional fact that macroscopic chirality 
violating sources for $J_5^{\mu}$ which would be sensitive to the 
anomaly are difficult to create. In QCD the situation is complicated by
the strong interactions in the infrared and chiral symmetry breaking.
The neutral member of the isotriplet of pseudoscalar Goldstone bosons 
in the low energy EFT is the $\pi^0$, whose decay to two photons, 
$\pi^0 \rightarrow 2 \gamma$ is correctly given by the triangle amplitude 
\cite{BarFriGM}. In fact, it was the experimental agreement between the 
measured decay rate to that predicted by the axial anomaly computed 
in the UV theory of $3$ colors of fractionally charged quarks that gave 
one of the strongest early confirmations of QCD. It is the fact that the 
anomaly may be computed in the UV theory but gives rise to a low
energy amplitude,  $\pi^0 \rightarrow 2 \gamma$ that led to the principle of 
anomaly {\it matching} \cite{tHooft}. 

The apparent massless pseudoscalar anomaly pole of in the {\it isosinglet} 
channel in the chiral limit of QCD is even more interesting. This pole mixes 
with the psuedoscalar axial gluon density $Q(x) = G^a_{\mu\nu}(x)\tilde G^{a \mu\nu}(x)$,
and gives rise to a non-vanishing susceptibility of axial gluon densities,
\be
\chi(k^2) = \int d^4x\,e^{ik\cdot x}\lag Q(x)\, Q(0)\rag\,,
\ee
as $k^2 \rightarrow 0$, despite the fact that $Q$ is a total derivative
and therefore one would naively expect $\chi(k^2)$ to be proportional to
$k^2$ and vanish in this limit. The fact that the susceptibility $\chi(0)$ is
non-vanishing is a direct effect of the massless anomaly pole \cite{Ven}.
The degree of freedom this infrared pole represents combines with a
non-dynamical but gauge invariant $Q^2$ term in the effective
action of QCD to yield finally one propagating massive isosinglet 
psuedoscalar state which can be identified with the $\eta'$ meson,
solving the $U(1)$ problem in QCD \cite{Ven}. Thus there is no
doubt that the pseudoscalar $0^-$ state which appears in the anomaly
channel in perturbation theory is physical and propagating in the final
S-matrix of the theory, but it becomes massive by a topological variant
of the Higgs mechanism \cite{ATT}.

The lesson to be taken away from this QCD example is that anomalies are a unique 
window which the low energy EFT provides to short distance physics. As such 
the anomalous Ward identities and the long distance effects they generate must 
be taken into account by explicitly adding the IR relevant terms 
they induce in the low energy effective action \cite{BarFriGM}.

\section{The Trace Anomaly in Two Dimensions}

Consider next gravity in $D=2$ dimensions. The local action,
\be
S_{cl}[g] = \int d^2x\, \sqrt{-g}\, (\gamma R - 2 \Lambda)
\label{Scltwo}
\ee
actually contains {\it no} local degrees of freedom at all, since all metrics in $D=2$
are locally conformally flat, and hence may be expressed in the form
\be
g_{\mu\nu} = \exp (2 \sigma) \, \bar g_{\mu\nu}
\label{confdef}
\ee
for some $\sigma (x)$ and a fixed metric $\bar g_{\mu\nu}$ which may be taken to be flat.
Owing to the identity,
\be
R \,\sqrt{-g} = \bar R \,\sqrt{-\bar g} - 2 \,\sqrt{-\bar g} \sqb \sigma\,,
\qquad (D=2)
\label{RRbar}
\ee
the $\sigma$ dependence of the Einstein-Hilbert term in (\ref{Scltwo}) is a
total derivative and gives no metric variation. Hence the theory described
by the local action (\ref{Scltwo}) has no local dynamical degrees of freedom
at all.

When massless conformal matter is coupled to the geometry, this situation
changes due to the conformal trace anomaly, \cite{BirDav}
\be
\lag T^{\mu}_{\ \mu} \rag = \frac{N}{24\pi}\, R\,,\qquad (D=2)
\label{trtwo}
\ee
where $N = N_S + N_F$ is the total number of massless fields, either
scalar ($N_S$) or fermionic ($N_F$). It is not difficult to show that
the amplitude $\int d^2 x\, e^{i k\cdot x} \,\lag T^{\mu\nu}(x) T^{\alpha\beta}(0)\rag$
at one-loop order develops a (gauge invariant) pole at $k^2 = 0$ when the mass 
of the quantum field  vanishes, whose residue is just proportional to the coefficient 
of the anomaly (\ref{trtwo})\cite{BerKohl}. Accordingly, no local coordinate invariant 
action exists whose metric variation leads to (\ref{trtwo}). 

A non-local action corresponding to (\ref{trtwo}) can easily be found by using the 
relation (\ref{RRbar}) in (\ref{trtwo}), which implies that the conformal variation, 
$\delta \Gamma/\delta \sigma = \sqrt{-g}\, \lag T^{\mu}_{\ \mu} \rag$ of the effective 
action $\Gamma$ reproducing the anomaly is linear in $\sigma$. Hence this 
Wess-Zumino effective action \cite{WZ} in two  dimensions is
\be
\Gamma_{WZ}^{(2)} [\bar g ; \sigma ] = \frac{N}{24\pi}  
\int\,d^2x\,\sqrt{-\bar g}
\left( - \sigma \sqb \sigma + \bar R\,\sigma\right)\,.
\label{WZact}
\ee
By solving (\ref{RRbar}) for $\sigma$ it is now straightforward to find a non-local 
scalar functional $S_{anom}[g]$ of  the full metric in (\ref{confdef}) such that
\be
\Gamma_{WZ}^{(2)} [\bar g ; \sigma ] = S_{anom}^{(2)}[g= e^{2\sigma}\bar g]
- S_{anom}^{(2)}[\bar g]\,,
\label{cohom}
\ee
upon also using the fact that $\sqrt{-g} \sq = \sqrt{-\bar g} \sqb$ is conformally invariant 
in two dimensions. In this way we find
\be
S_{anom}^{(2)}[g] = -\frac{N}{96\pi} \int\,d^2x\,\sqrt{-g}
\int\,d^2x'\,\sqrt{-g'}\, R(x)\,{\sq}^{-1}(x,x')\,R(x')\,,
\label{acttwo}
\ee
with ${\sq}^{-1}(x,x')$ denoting the Green's function inverse of the scalar 
differential operator $\sq$. The parameter $N$ is replaced by $N-25$ if account is 
taken of the contribution to the anomaly coefficient of the metric fluctuations themselves 
in addition to those of the $N$ matter fields.

The anomalous effective action (\ref{acttwo}) is a scalar under coordinate 
transformations and therefore fully covariant and geometric in character,
as required by the Equivalence Principle. However since it involves the 
Green's function $\sq^{-1}(x,x')$, which requires boundary conditions for 
its unique specification, it is quite non-local, and dependent upon more 
than just the local curvature invariants of spacetime. In this important 
respect it is quite different from the classical action (\ref{Scltwo}),
and describes rather different physics. In order to expose that physics 
it is most convenient to recast the non-local and non-single valued 
functional of the metric, $S_{anom}^{(2)}$ into a local form by 
introducing a scalar auxiliary field $\varphi$ satisfying \cite{Rie}
\be
- \sq \varphi = R\,.
\label{auxeomtwo} 
\ee
Then one may check that varying
\be
S_{anom}^{(2)}[g;\varphi]  \equiv -\frac{N}{96\pi} \int\,d^2x\,\sqrt{-g}\,
\left(g^{\mu\nu}\,\nabla_{\mu} \varphi\,\nabla_{\nu} \varphi - 2 R\,\varphi\right)
\label{actauxtwo}
\ee
with respect to $\varphi$ gives the eq. of motion (\ref{auxeomtwo})
for the auxiliary field, which when solved formally by $\varphi =-{\sq}^{-1}R$
and substituted back into $S_{anom}^{(2)}[g;\varphi]$ returns the non-local
form of the anomalous action (\ref{acttwo}), up to a surface term. 
The non-local information in addition to the local geometry which was previously 
contained in the specification of the Green's function ${\sq}^{-1}(x,x')$ 
now resides in the local auxiliary field $\varphi (x)$, and the freedom 
to add to it homogeneous solutions of (\ref{auxeomtwo}).

In the local form (\ref{actauxtwo}), we see that a new local scalar degree of freedom has
appeared in the form of the auxiliary field $\varphi$, which was not present in the
original classical action (\ref{Scltwo}). This field is associated with the scalar conformal
deformations of the metric, which now fluctuates freely thanks to the kinetic term
in (\ref{actauxtwo}), whereas the metric was constrained in (\ref{Scltwo}). The massless
pole in the intermediate state of $\int d^2 x\, e^{i k\cdot x}\, \lag T^{\mu\nu}(x) T^{\alpha\beta}(0)\rag$
is exactly the $\varphi$ propagator. Its fluctuations lead to the gravitational
``dressing" of the critical exponents of conformal matter in a gravitational background \cite{KPZ}.

Since critical exponents are characteristic of long range fluctuations in a second
order phase transition, this shows that the effective action of the anomaly  (\ref{actauxtwo})
is definitely a relevant operator in the infrared. The anomalous action is also
responsible for the infrared {\it running} of $\Lambda$. In other words, the bare
parameter of the classical action is renormalized by the quantum fluctuations of
the $\varphi$ field, and becomes scale dependent. For all of these reasons
the action  (\ref{actauxtwo}) and the additional scalar degree of freedom
contained in it must be added to the classical action (\ref{Scltwo}), 
to get a complete low energy EFT of two-dimensional gravity. 

\section{Massless Scalar Degrees of Freedom in $4D$ Gravity}
 
With the examples of the QED/QCD axial anomaly and the conformal anomaly 
in two dimensions, we consider finally
the conformal or trace anomaly in $D=4$ dimensions, 
\be
\langle T^{\mu}_{\ \mu} \rangle
= b F + b' \left(E - \frac{2}{3}\sq R\right) + b'' \sq R + \sum_i c_iH_i\,,
\label{tranomgen}
\ee
where
\bes\bea
&&E \equiv ^*\hskip-.2cmR_{\mu\nu\alpha\beta}\,^*\hskip-.1cm R^{\mu\nu\alpha\beta} = 
R_{\mu\nu\alpha\beta}R^{\mu\nu\alpha\beta}-4R_{\mu\nu}R^{\mu\nu} + R^2 
\,, \label{EFdef}\\
&&F \equiv C_{\mu\nu\alpha\beta}C^{\mu\nu\alpha\beta} =
 R_{\mu\nu\alpha\beta}R^{\mu\nu\alpha\beta}
-2 R_{\mu\nu}R^{\mu\nu}  + \frac{R^2}{3}\,.
\eea\ees
and $H_i$ denotes any number of dimension $4$ conformally invariant scalars
constructed from the gauge fields externally coupled to the matter in question. 
For example in QED in {\it flat} space with $E=F=0$, with massless fermions 
coupled to electromagnetism,  $H = F_{\mu\nu}F^{\mu\nu}$ and one finds 
\be
\lag T^{\mu}_{\ \mu}\rag_{_A} \Big\vert_{m=0,flat} = c\, H =
-\frac{e^2}{24\pi^2}\, F_{\mu\nu}F^{\mu\nu}\,,
\label{tranom}
\ee
in complete analogy with (\ref{axanom}). The coefficients $b$ and $b'$ in 
(\ref{tranomgen}), like the coefficient in (\ref{axanom}) 
do not depend on any ultraviolet short distance cutoff, but instead
are determined only by the number and spin of massless fields 
\cite{BirDav,CapDuffDes}, 
\bes\bea
b &=& \frac{\hbar}{120 (4 \pi)^2}\, (N_S + 6 N_F + 12 N_V)\,,\\
b'&=& -\frac{\hbar}{360 (4 \pi)^2}\, (N_S + \frac{11}{2}N_F + 62 N_V)\,,
\label{bprime}
\eea \label{bbprime}\ees
\noindent with $(N_S, N_F, N_V)$ the number of fields of spin 
$(0, \frac{1}{2}, 1)$ respectively.

For the flat space trace anomaly in an external electromagnetic field (\ref{tranom}) 
one can perform a full analysis of the one-loop triangle contribution to the amplitude,
\be
\Gamma^{\mu\nu\alpha\beta} (p,q) = \int\,d^4x\,\int\,d^4y\ e^{ip\cdot x + i q\cdot y}\,
= \frac{ \delta^2 \lag T^{\mu\nu} (0) \rag_A}
{\delta A_{\alpha}(x)\delta A_{\beta}(y)} \bigg\vert_{A=0}\,,
\label{TJJ}
\ee
analogous to (\ref{GJJJ}). This triangle amplitude analogous to that in Fig. 1 (with $J^{\mu}_5$
replaced by $T^{\mu\nu}$) also develops a pole at $k^2 = 0$
when the combined limits $p^2 = q^2 = m^2 = 0$ are taken. The $0^+$ scalar pole
survives in matrix elements of the stress-energy tensor of massless fermions
to physical two-photon states, analogous to (\ref{matpol}). The residue of the pole 
is proportional to the coefficient of the anomaly. The discontinuity of the $\lag TJJ \rag$ 
triangle diagram also exhibits a $\delta$ function in the intermediate two-electron state 
with total spin zero, which signifies that a new massless scalar degree of freedom 
with gravitational coupling is required by the anomaly \cite{GiaMot}. 

The effective action of the massless degree(s) of freedom is again non-local
in terms of the original local metric and electromagnetic field strengths.
However it can be rewritten in a completely local form by the introduction
of the {\it two} local scalar fields whose quanta are responsible for the massless poles
in anomalous amplitudes such as the $\lag TJJ \rag$ and $\lag TTT \rag$ triangle 
diagrams. By following steps analogous to those in the two dimensional case of
the previous section, explicitly integrating the anomaly eq. (\ref{tranomgen}) and 
introducing two auxiliary fields $\varphi$ and $\psi$ to account for the two independent 
invariants $b'E$ and $bF$ in the trace anomaly (\ref{tranomgen}) one finds \cite{Rie,FraTse,Shap,MotVau}
\bea
&& S_{anom}[g;\varphi,\psi] =\frac{b'}{2}\,\int\,d^4x\,\sqrt{-g}\ 
\left\{ -\varphi \Delta_4 \varphi + \left(E - \frac{2}{3} \sq R\right) \varphi \right\}
\nn\\
&&+ \frac{b}{2} \,\int\,d^4x\,\sqrt{-g}\ \left\{ -2\varphi \Delta_4 \psi + 
\left( F + \frac{c}{b}\, H\right)\,\varphi 
+ \left(E - \frac{2}{3} \sq R\right) \psi \right\}
\label{locaux}
\eea
where in QED, $c = -e^2/24\pi^2$, $H = F_{\mu\nu}F^{\mu\nu}$ and in general,
\be
\Delta_4 \equiv \sq^2 + 2 R^{\mu\nu}\nabla_{\mu}\nabla_{\nu} - \frac{2}{3} R \sq + 
\frac{1}{3} (\nabla^{\mu} R)\nabla_{\mu} \,.
\label{Deldef}
\ee
By variation of (\ref{locaux}) the auxiliary scalar fields satisfy the linear fourth order eqs. of motion,
\bes\bea
&& \Delta_4\, \varphi = \frac{1}{2}E - \frac{1}{3} \sq R\,,     \label{eomphi}\\
&& \Delta_4\, \psi = \frac{1}{2} F  + \frac{c}{2b} \,H\,.
\eea \label{auxeom}\ees
The effective action (\ref{locaux}) analogous to (\ref{actauxtwo}) in $D=2$ generates all the 
anomalous amplitudes, by successive variations with respect to background 
metric and/or external gauge potentials, including diagrams with multiple 
stress-energy tensor insertions, such as $\lag TTT...JJ\rag$
and $\lag TTT ...\rag$.

Since from the free variation of the effective action (\ref{locaux}) $\varphi$ and $\psi$ obey 
non-trivial massless wave eqs., they are additional massless scalar degrees of freedom in
low energy gravity, over and above the usual transverse, tracefree gravitational waves
of the Einstein theory, and can have long range, macroscopic effects. The
poles in the amplitude (\ref{TJJ}) survive in low energy scattering processes 
involving two photons with a gravitational strength \cite{GiaMot}. 

For QED the effects of the trace anomaly $\lag TJJ\rag$ in flat space are again screened 
at distances greater than $\hbar/mc$ by the finite mass of the electron. However the trace 
anomalies in $ \lag TTT...\rag $ amplitudes which couple to purely gravitational sources and
scattering processes are mediated by truly massless particles, such as the photon itself.
Unlike the axial case, where non-trivial sources for $J_5^{\mu}$ may not be easy to
come by, the sources here are gravitational which are omnipresent in the Universe.
Since the coupling is through the stress-energy, it is universal but weak. However
gravity is unscreened by any other interaction and its effects are cumulative over
large distances. The long range nature of massless fluctuations make the scalar degrees 
of freedom  contained in the effective action (\ref{locaux}) relevant at the even
the very largest macroscopic distance scales of cosmology. 

\section{Dynamical Vacuum Energy}

The example of the axial anomaly, well-tested in QCD, and the logarithmic scaling of the
anomaly action (\ref{locaux}) imply that the full effective action of low energy gravity
should be the sum of the classical Einstein-Hilbert action of classical General Relativity
together with the effective action (\ref{locaux}). In this EFT the additional long range scalar 
modes have some interesting effects. 

With a covariant action functional (\ref{locaux}) one may compute a covariantly conserved
stress-energy tensor \cite{MotVau}
\be
T_{\mu\nu}^{(anom)} \equiv  - \frac{2}{\sqrt{-g}} \frac{\delta S_{anom}}{\delta g^{\mu\nu}}\,
\label{Tanom}
\ee
the $b'$ term of which gives the tensor,
\bea
E_{\mu\nu} &=& - \frac{2}{3}\partial_\mu\partial_\nu \sq \varphi  
-2(\partial_{(\mu}\varphi) \partial_{\nu)} \sq \varphi
+ 2(\partial_\mu\partial_\nu\varphi)\sq \varphi
+ \frac{2}{3}(\partial_\alpha \varphi)(\partial^\alpha\partial_\mu\partial_\nu\varphi)\nn\\
&&\hspace{-3mm} -\frac{4}{3}\,(\partial_\mu\partial_\alpha\varphi)
(\partial_\nu\partial^\alpha\varphi) + \frac{1}{6}\, \eta_{\mu\nu}\, \left\{-3\, (\sq\varphi)^2 
+ \sq \left[(\partial_\alpha\varphi)(\partial^\alpha\varphi)\right] \right\}
\label{Eab}
\eea
in flat space, which is conserved by use of the eqs. of motion (\ref{auxeom}).
A particular solution of (\ref{eomphi}) in flat space with $E= \sq R=0$ is
\be
\varphi = a\, \frac{z^2}{d^2}\,,
\label{phiz}
\ee
(with the second auxiliary field $\psi = 0$) which leads to
\be
T_{\mu\nu}^{(anom)} = b' E_{\mu\nu} = -\frac{2 b'a^2}{3 d^4} \,\left(\eta_{\mu\nu} 
- 4 \delta_{\mu}^z\delta_{\nu}^z\right)= \frac{C}{d^4}\, {\rm diag} (-1,1,1, -3)\,.
\label{TCas}
\ee
This is exactly the form of the Casimir vacuum stress-energy between two 
infinite parallel conducting plates a distance $d$ apart in (\ref{Casimir}). 
The constant $a$ and therefore $C = -2b' a^2/3$ depends upon the boundary 
conditions imposed on the conductors. Notice from (\ref{bprime}) that $b' <0$ 
so that the constant $C > 0$, corresponding to an attractive force between the 
plates for any real $a$. It is remarkable that the auxiliary field $\varphi$ and 
its stress tensor, obtained from the local form of {\it quantum} anomaly in the 
{\it trace} of the stress-energy tensor in {\it curved} space may be regarded 
through a particular homogeneous solution (\ref{phiz}) of the {\it classical} 
eqs. (\ref{auxeom}) as responsible for the Casimir stress tensor (\ref{TCas}) 
in {\it flat} space, where it is {\it tracefree}.

Several other examples of the auxiliary fields and stress tensor (\ref{Tanom}) 
in curved black hole and cosmological spacetimes have now been studied 
and used to compute vacuum polarization, illuminating their physical 
meaning \cite{MotVau,RN}. Solving classical differential eqs. for the auxiliary 
fields in a given fixed spacetime background allows one to survey a great number 
of physical states of the underlying quantum field theory, taking account the spin of 
the field through the $b, b'$ coefficients (\ref{bbprime}), rather than having to 
decompose the solutions of the field eqs. for each spin into normal modes, impose 
boundary conditions on those modes, construct the stress tensor and regularize it 
and renormalize it in each quantum state. The stress-energy (\ref{Tanom}) is 
particularly important in the vicinity of black hole and cosmological horizons, 
which it can dominate even the classical curvature terms, and lead to large 
quantum vacuum polarization effects there \cite{MotVau,RN,AndMolMot}.

In addition, when the scalar fields $\varphi, \psi$ in (\ref{locaux}) are treated as dynamical
fields and quantized in their own right, they lead to infrared renormalization and finite
volume dependence of the effective cosmological term (\ref{Lnum}).
In other words the quantity $\lambda = \hbar G \Lambda/c^3$ becomes a dynamical 
quantity in the EFT, running with IR renormalization scale as every other coupling 
affected by light degrees of freedom. A one-loop calculation with the auxiliary field 
propagator given by the inverse of (\ref{Deldef}) gives the volume scaling relation \cite{AntMazMot}
\be
V \frac{d\lambda}{d V}  = 4\, (2 \delta - 1)\, \lambda\,,
\label{renL}
\ee
with the anomalous dimension,
\be
2 \delta - 1 = 
\frac{\sqrt{1 - \frac{8}{Q^2}} - \sqrt{1 - \frac{4}{Q^2}}}
{1 + \sqrt{1 - \frac{4}{Q^2}}} \le 0\,, 
\label{scal}
\ee
and
\be
Q^2 \equiv -32 \pi^2 b' = \frac{1}{180}\, \left(N_S + \frac{11}{2} N_F + 62 N_V\right) + Q^2_{grav}
\label{Qdef}
\ee
in terms of the anomaly coefficient $b'$ in (\ref{tranomgen}), and $Q^2_{grav}$, 
the contribution of graviton fluctuations to the anomaly coefficient
(approximately $7.9$). The anomalous scaling dimension (\ref{scal}) is negative for all 
$Q^2 \ge 8$. This implies that the dimensionless cosmological term $\lambda$ 
has an infrared fixed point at zero as the volume $V\rightarrow \infty$. Thus the cosmological 
term is dynamically driven to zero as $V\rightarrow \infty$ by infrared fluctuations 
of the conformal part of the metric described by (\ref{locaux}). There is no fine tuning 
involved here and no free parameters enter except $Q^2$, which is determined 
by the trace anomaly coefficient $b'$ by (\ref{Qdef}). Once $Q^2$ is assumed
to be positive, then $2 \delta - 1$ is negative, and $\lambda$ is driven to zero 
at large volumes or large distances by the conformal fluctuations of the metric.
This identifies a mechanism for the dynamical screening of the vacuum energy 
at large distances, relying only on the four dimensional quantum
physics of the trace anomaly with no additional assumptions.

Thus, the fluctuations of the new scalar degrees of freedom in the effective action
(\ref{locaux}) of the anomaly allow the cosmological ``constant" vacuum energy
of classical General Relativity to vary dynamically. This is
qualitatively similar to the effect of the $\varphi$ conformal degree of freedom
in $2D$ gravity (\ref{actauxtwo}) at second order critical points \cite{KPZ}. The fixed
point of $\lambda = 0$ is stable to marginal deformations by the Einstein-Hilbert
terms, and describes a quantum conformal phase of $4D$ gravity. To take account
of this mechanism and understand the role of the conformal phase where the
fluctuations of the new dynamical scalar degrees of freedom are important
in a consistent cosmological theory of vacuum dark energy is the remaining task.

\section{Linear Response in de Sitter Space and Cosmological Horizon Modes}

To this end we have recently studied the effect of the scalar fluctuations in the
linear response of coupled matter-geometry perturbations around de Sitter
spacetime, relevant for both inflationary and present day dark energy cosmology
\cite{AndMolMot}. Here only the main results are summarized.

Linear response in gravity means solving the linear eqs. for small perturbations of the
metric and matter stress-energy renormalized expectation value $\lag T^{\mu}_{\ \nu}\rag_{_R}$ 
expanded around a self-consistent solution of the semi-classical Einstein equations.
A self-consistent solution of these eqs. is de Sitter space,
\be
ds^2 = -c^2 d\tau^2 + a^2(\tau)(dx^2 + dy^2 + dz^2) = -c^2 d\tau^2 +  e^{2H\tau} d \vec x \cdot d \vec x
\label{dSRW}
\ee
with $a(\tau) = e^{H\tau}$ the de Sitter Robertson-Walker (RW) scale factor,
and conformal matter fields in their de Sitter invariant Bunch-Davies state \cite{BirDav}. 
The linear variation of the semi-classical Einstein eqs. is:
\be \delta
\left\{R^{\mu}_{\ \nu} - \frac{R}{2} \delta^{\mu}_{\ \nu} + \Lambda
\delta^{\mu}_{\ \nu}\right\} = \frac{8\pi G}{c^4}\, \delta\, \lag T^{\mu}_{\ \nu}\rag_{_R}\, ,
\label{linresgen}
\ee
The variation on the left side is purely geometrical, obtained by varying the
metric from its de Sitter value $g_{\mu\nu}$ given in RW coordinates (\ref{dSRW}) 
to $g_{\mu\nu} + \delta g_{\mu\nu}$. The variation of the expectation value on the 
right side contains two kinds of terms. The first kind are also proportional to the 
metric variation $h_{\mu\nu} =\delta g_{\mu\nu}$ and involve the retarded 
response function of stress-energy fluctuations, namely 
$\theta(t-t') \lag [T^{\mu}_{\ \nu}(x), T^{\alpha\beta}(x')] \rag$ integrated
over all points $x'$ in the causal past of the point $x$. It turns out that this first kind of 
variation leads to solutions of linear response eqs. (\ref{linresgen}) which have spacetime
dependence only on the Planck scale (\ref{Lpl}). Since this is the ultraviolet cutoff scale 
at which the semi-classical EFT breaks down, one cannot trust any physical conclusion 
obtained with this first kind of solution of (\ref{linresgen}).

The second kind of term in the stress-energy variation $\delta\, \lag T^{\mu}_{\ \nu}\rag_{_R}$ 
in (\ref{linresgen}) arise from the possibility of varying the quantum state of the field in 
which the expectation value $\lag T^{\mu}_{\ \nu}\rag_{_R}$ is evaluated,
independently of the variation of the metric. The quantum state is specified by boundary 
conditions on the cosmological horizon scale $c/H$, having nothing to do with the 
microscopic Planck scale $L_{Pl}$. The scalar auxiliary field effective action (\ref{locaux}) 
and eqs. (\ref{auxeom}) in de Sitter space parametrize additional state dependent contributions
to $\delta\, \lag T^{\mu}_{\ \nu}\rag_{_R}$ from the variation of $\delta T_{\mu\nu}^{(anom)}$ 
of (\ref{Tanom}). The gauge invariant combination of auxiliary field $\delta\varphi$ and metric 
perturbation $h_{\tau\tau}$ given by 
\be
u = \frac{1}{H^2}\left(\frac{\partial^2}{\partial\tau^2}  + H \frac{\partial}{\partial \tau} - 
\frac{\vec \nabla^2}{a^2}\right) \delta\varphi - 2\, h_{\tau\tau}
\ee
in the gauge $g^{ij}h_{ij} = 0  = \nabla^ih_{i\tau} - \frac{1}{2} h_{\tau\tau}$ satisfies 
the {\it second} order homogeneous eq., \cite{AndMolMot}
\be
\left( \frac{\partial^2}{\partial\tau^2}+ 5H \frac{\partial}{\partial\tau}+ 6H^2 
- \frac{\vec\nabla^2}{a^2} \right)u  = 0\,,
\label{uveom}
\ee
the general solution of which is a linear combination of
\be
u_{\vec k ,\pm}(\tau, \vec x)  =  v_{\vec k ,\pm}(\tau, \vec x)  = 
\frac{1}{a^2} \,\exp \left ( \pm \frac{ik}{Ha}  + i \vec{k} \cdot \vec x\right)\,.
\label{uvsolns}
\ee
in Fourier space. These modes and those of the same form arising from the second
auxiliary field $\psi$ (called $v$), give rise in the linear response eq. (\ref{linresgen}) to
perturbations of the Ricci tensor,
\be
\delta R^{\tau}_{\ \tau}  = - \delta R^i_{\ i} = 
\frac{\varepsilon' }{2a^2}\, \stackrel{\rightarrow}{\nabla}\!\!^2\,
u- \frac{\varepsilon }{6a^2}\,
\stackrel{\rightarrow}{\nabla}\!\!^2\, v\,,
\label{tteq}
\ee
with $\delta R = 0$ and
\bes
\bea
\varepsilon \equiv && 32 \pi GH^2 b \,,\\
\varepsilon' \equiv && -\frac{32\pi }{3} G H^2 b'\,.
\eea
\label{pert}
\ees

\vspace{-3mm}
\noindent
Thus the auxiliary fields of the anomaly action yield non-trivial
gauge invariant solutions for the stress tensor and corresponding
linearized Ricci tensor perturbations (\ref{tteq}). Being solutions of
(\ref{uveom}) which itself is independent of the Planck scale, these solutions
vary instead on arbitrary scales determined by the wavevector $\vec k$, and are
therefore genuine low energy modes of the semi-classical effective theory. The
Newtonian gravitational constant $G$ and the Planck scale enter
(\ref{tteq}) only through the small coupling parameters $\varepsilon$
and $\varepsilon'$ between the auxiliary fields and the metric perturbation.
In the limit of either flat space, or arbitrarily weak coupling
$GH^2 \rightarrow 0$ these modes decouple from the metric
perturbations at linear order. Thus, there is no problem with possible
negative metric modes of the $\varphi,\psi$ fields propagating to infinity 
and leading to a non-unitary $S$-matrix in flat space. However, in a
curved space such as de Sitter space these modes can have physical
infrared effects at the horizon scale where they do couple.

The infrared scalar $u$ and $v$ modes are associated with the cosmological
horizon scale $c/H$ in de Sitter space and for that reason may be called 
{\it cosmological horizon modes}. To show this connection, one may introduce the
static coordinates of de Sitter space, {\it viz.}
\be
ds^2 = - (c^2-H^2r^2)dt^2 + \frac{dr^2}{1-H^2r^2/c^2}
+ r^2 (d\theta^2 + \sin^2\theta d\phi^2)\,,
\label{static}
\ee
related to the RW coordinates (\ref{dSRW}) by the coordinate transformation,
\bes
\bea
&& r = |\vec x |\, e^{H \tau} \,,\\
&& t = \tau - \frac{1}{2H} \ln \left(1 - \frac{H^2 |\vec x|^2}{c^2} e ^{2H\tau}\right)\,.
\eea
\ees
In these static coordinates the eqs. of motion (\ref{uveom}) possess the
time independent solution,
\be
u = v = \frac{1}{1- H^2r^2/c^2}
\ee
which diverge on the cosmological horizon $r= c/H$ centered at the origin.
The corresponding stress-energy tensor perturbation in the static frame is
\be
\delta \lag T^r_{\ r}\rag_{_R} = \delta \lag T^{\theta}_{\ \theta}\rag_{_R} =
\delta \lag T^{\phi}_{\ \phi}\rag_{_R} = -\frac{1}{3}\, \delta \lag T^t_{\ t}\rag_{_R} =  
 C\, \frac{\hbar\, H^4/c^3}{(1-H^2r^2/c^2)^2}\,,
 \label{Thoriz}
\ee
with a quadratically {\it divergent} value on the cosmological horizon.
This form of the stress tensor perturbation is also the form of a finite
temperature fluctuation away from the Hawking-de Sitter temperature 
$T_{_H} = \hbar H/2\pi k_{_B}$ of the Bunch-Davies state in static coordinates \cite{AndHisSam}. 
It corresponds therefore to a change of the boundary conditions on the state 
of the underlying quantum fields on the horizon, with a corresponding change
in the vacuum polarization effects of the fields near the horizon. 

That these Casimir like effects are dynamical and functions of the boundary conditions
is not surprising given our experience with the Casimir effect in flat space. It is
also known that the Casimir stress-energy can diverge as a curved surface of
a perfect conductor is approached \cite{DeuCan}. In electromagnetism we know that
perfect conductors do not exist and this mathematical divergence is cut off by the
finite conductivity and skin depth of a metallic surface \cite{Cas}. A diverging stress 
tensor on the cosmological horizon signals the breakdown of boundary conditions there
as well, except that in classical General Relativity the horizon is supposed to be
purely a mathematical boundary with no physical stress tensor. Large stresses
on the horizon of the form (\ref{Thoriz}) suggest that this assumption may be incorrect.

In the semi-classical EFT with $S_{anom}$ included in the effective action there are 
additional scalar degrees of freedom in gravity that become important in the vicinity 
of geometries with horizons and there their fluctuations must be taken into account, just
as charged matter fluctuations must be taken into account in an imperfect conductor. 
The stress-energy (\ref{Thoriz}) of these degrees of freedom in the vicinity of the 
cosmological horizon can cause large backreaction on the classical geometry.
Because the horizon is a null surface, the massless propagator pole associated with
the anomalous amplitudes as in (\ref{matpol}) can lead to large quantum correlations
in multi-stress tensor amplitudes $\lag TTT...\rag$ on the horizon as well. Such large 
amplitudes in fluctuations from the mean $\lag T^{\mu}_{\ \nu}\rag$ is characteristic of a
{\it phase transition} in which the semi-classical mean field theory breaks down,
and where the rigid cosmological constant term of the classical theory can change. 

Thus the fluctuations of the scalar degrees of freedom determined by the anomaly 
may lead to a phase transition to precisely the conformally invariant phase of 
gravity described by the fixed point $\lambda = 0$ of (\ref{renL}) in the near vicinity of 
the horizon. This suggests a rather different cosmological model than the standard one,
in which we live inside a kind of ``bubble" of vacuum energy condensate, with a 
preferred origin and a physical surface at the cosmological horizon \cite{AntMazMot,gstar}. 
At the horizon the quantum fluctuations of the scalar degrees of freedom contained
in the anomaly lead to a phase transition in which the spacetime condensate $\Lambda$
``melts." The $\Lambda$ condensate will then behave very much like the gluon 
condensate in the QCD bag model of hadrons.

In the low energy EFT of gravity that takes account of the trace anomaly,
the value of the cosmological dark energy in the interior is dynamical and should be
fixed by the boundary conditions on the surface at the infrared Hubble scale $c/H$, 
much as the Casimir effect is, with no regard to the ultraviolet divergent and clearly
incorrect estimate of (\ref{zeropt}). The consequences of this dynamical dependence
upon infrared boundary conditions of the vacuum energy for cosmology, microwave
background anisotropies and non-Gaussianities, and in the gravitational
collapse problem are presently under investigation.

\end{document}